\documentclass[
aps
,nofootinbib
,showkeys
,showpacs
,amsfonts
,eqsecnum
,nofootinbib
]{revtex4%
}

\usepackage{epsfig}
\usepackage{graphicx}
\usepackage{bm}
\usepackage{amsmath}
\usepackage{amsfonts}
\usepackage{amssymb}

\newcommand{\tr}{{\rm tr \,}}

\newcommand{\hrho}{{\hat{\rho}}}
\newcommand{\alphaa}{a}
\newcommand{\betaa}{b}

\newcommand{\eq}[1]{eq.~(\ref{eq:#1})}

\newcommand{\ignore}[1]{}

\newcommand{\vx}{\bm x}
\newcommand{\vk}{\bm k}
\newcommand{\vn}{\bm n}
\newcommand{\vJ}{\bm J}
\newcommand{\vL}{\bm L}
\newcommand{\vS}{\bm S}
\newcommand{\vomega}{\bm \omega}
\newcommand{\vsigma}{\bm \sigma}
\newcommand{\vnabla}{\bm \nabla}

\begin{document}

\title{Statistical consistency of quantum-classical hybrids}

\author{L. L. Salcedo}
\affiliation{Departamento de F{\'\i}sica At\'omica, Molecular y Nuclear, and
Instituto Carlos I de F{\'\i}sica Te\'orica y Computacional,
  Universidad de Granada, E-18071 Granada, Spain}
%\email{salcedo@ugr.es}

\date{\today}

\begin{abstract}
After formulating a no-go theorem for perfect quantum-classical hybrid
systems, a new consistency requirement based on standard statistical
considerations is noted. It is shown that such requirement is not fulfilled by
the mean-field approach, nor by the statistical ensemble approach. Further
unusual features of the latter scheme are pointed out.
\end{abstract}



\maketitle
\tableofcontents

%\newpage

\section{Introduction}
\label{sec:1}

The study of the interrelation between classical and quantum dynamics is as
old as the quantum theory itself. The Copenhagen interpretation invokes a
classical measuring device in interaction with the quantum system to be
measured \cite{Borh:1958,Heisenberg:1958,dEspagnat:1976,Allahverdyan:2011cx}.
However, while classical or quantum dynamics are each one internally
consistent by themselves and formally similar, their coupling is not
straightforward, and in fact poses a problem of consistency
\cite{DeWitt:1962bu,Boucher:1988ua,Gisin:1990,Salcedo:1994sn,Salcedo:1996jr,%
  Halliwell:1997hy,Caro:1998us,Peres:2001,Salcedo:2003mp,Terno:2004ti,%
  Salcedo:2007,Agostini:2007}.  The most immediate approach is the mean-field
scheme (or semiclassical or Ehrenfest method), in which the classical system
couples to the quantum one through the expectation values. This scheme is
robust and elegant but not realistic as it misses the back-reaction from the
quantum fluctuations \cite{Hu:1994iw,Wald:1994bk,Craig:2005}. For instance, in
cosmology it leads to problems to account for the local anisotropy in the
early universe \cite{Brandenberger:1984cz}. Many other coupling schemes for
hybrid classical-quantum systems have been proposed in the literature
\cite{Delos:1972,Sudarshan:1976bt,Sherry:1978ea,Aleksandrov:1981,%
  Boucher:1988ua,Jones:1993,Salcedo:1994sn,Anderson:1995tn,Diosi:1995qs,%
  Prezhdo:1996gs,Diosi:1999py,Gindensperger:2000,Peres:2001,Dias:2001,
  Nielsen:2001wh,Prezhdo:2001,Kisil:2005,Sergi:2005,Prezhdo:2006,Elze:2011hi}.
As shown in \cite{DeWitt:1962bu} (of course, under pertinent assumptions) as
soon as a classical degree of freedom couples to a quantum system, a
consistent description is only possible if the classical system inherits
fluctuations which turn it quantal. This type of argument, as well as the
accuracy of quantum mechanics in every prediction, would point to the
conclusion that only quantum systems exist in nature. While this view is quite
extended among the scientific community it is by no means universally
accepted, due to the conceptual problems involved in the quantum description
\cite{Einstein:1935,Bohm:1966,Bell:1987hh,Schmelzer:2011}.  Also the technical
and conceptual difficulties encountered in quantum gravity have prompted a
certain discussion regarding the necessity or not of treating gravity quantum
mechanically
\cite{Moller:1952,Rosenfeld:1963,Kibble:1979jn,Kibble:1980ia,Isham:1980sb,%
  Page:1981aj,Alvarez:1988tb,Hu:2002jm,Boughn:2008jx,Carlip:2008zf,%
  Mattingly:2009,Diosi:2011vu}.

In our view the problem of ``semiquantization'', that is, treating
simultaneously classical and quantum observables in interaction and
constructing a consistent algebra of such observables, is closely related to
the problem of ``quantization'', i.e., how to obtain the quantum version of a
system and its algebra of observables, from its classical version. For the
latter there are well known no-go theorems
\cite{Groenewold:1946,VanHove:1951,Abraham:1978bk} and similar negative
theorems have been forwarded for the semiquantization problem
\cite{Salcedo:1996jr,Caro:1998us,Peres:2001,Salcedo:2003mp,Terno:2004ti,%
  Salcedo:2007,Agostini:2007}. If all physical systems are described by
quantum mechanics, both problems would be of technical type, rather than of
fundamental type. The quantization program would be to use the classical
description as guidance to construct the correct quantum description among the
several possible ones with the same classical limit. The semiquantization
program would be how to construct a useful approximate description in which
some sector of the theory is treated classically to make the treatment
accessible to computation. This is actually the situation in many applications
in physics and quantum chemistry
\cite{Delos:1972,Delos:1981,Halcomb:1986,Hu:1994iw,Craig:2005,%
  Micha:2007,Bousquet:2011}.

One goal of the present work is to reanalyze the conditions fulfilled by
classical and by quantum systems, conditions which guarantee a consistent
description. In \cite{Salcedo:1996jr,Caro:1998us} it was already shown that a
perfect quantum-classical hybrid, i.e., sharing all the nice properties of
classical and quantum systems is not viable (at least without restrictions on
the allowed interactions). A technical limitation was that variables of the
position-momentum type were assumed in the quantum sector.\footnote{This is an
  important limitation but the proof still covers all degrees of freedom of
  bosonic type, since the position-momentum pair can be combined to form
  bosonic creation and annihilation operators and the corresponding bosonic
  quantum fields.} More general proofs have been obtained in
\cite{Sahoo:2004,Dass:2009qu}. Here we present an extremely streamlined proof
of this no-go theorem. To have a proof as simple as possible should be of
interest for modelers of quantum-classical hybrid dynamics. The failure of
perfect hybridization has some consequences if quantum-classical hybrids are
taken at the fundamental level, i.e., not as approximations. If such hybrid
system would exist in nature they would have emergent properties
\cite{Carroll:2010zza} not shared by the purely classical or the purely
quantum dynamics. This is by itself problematic because the prime example of
classical-quantum interacting system, according to the Copenhagen
interpretation, would be a quantum measurement. No emergent phenomena has been
detected there. Easy or hard to understand from a conceptual point of view,
what is seen there is part of the standard quantum mechanics.

However, the main new contribution of this work is the introduction of a
further consistency requirement to be fulfilled by quantum-classical hybrid
dynamics. Essentially, we note that mixed statistical quantum states, a
density matrix, can be decomposed in many different ways into pure states.
Likewise, a mixed statistical classical state can be decomposed as statistical
combinations of other mixed classical states. For either purely classical or
purely quantum systems, the concrete decomposition is not relevant, only the
statistical mixture is. As it turns out, it is a non trivial requirement that
the evolution of quantum-classical hybrid systems should be independent of the
concrete decomposition. This requirement was briefly touched upon in
\cite{Salcedo:2003mp}. Here we analyze it in more detail and show that the
mean-field approach and the statistical ensemble scheme in configuration space
\cite{Hall:2005ax,Hall:2008,Reginatto:2009,Chua:2011fz} fail to fulfill the
requirement. So, for instance, if the violation of this principle were an
emergent feature of the hybrid systems, it would be possible to say which is
the ``true'' polarization of the electrons in an unpolarized electron
beam. This provides a test for such hybrid schemes. In passing, the hybrid
scheme based on statistical ensembles in configuration space is analyzed in
some detail. Further emergent features are unveiled for this approach, such as
ghost coupling between non interacting (but entangled) classical and quantum
sectors, and non conservation of angular momentum in presence of spin, or of
internal symmetries in general.

In Section \ref{sec:2} the proof of the no-go theorem for perfect
quantum-classical hybrids is presented. In Section \ref{sec:3} the statistical
consistency requirement is introduced and applied to the mean-field
approach. In Section \ref{sec:4} the statistical ensemble approach is
analyzed, in particular regarding its statistical consistency. In Section
\ref{sec:5} we present our conclusions.

\section{No-go theorem}
\label{sec:2}

A classical system with $n_c$ degrees of freedom can be described using the
canonical formalism, i.e., by means of the phase space variables position and
momentum, $x_i,k_i$, ~$i=1,\ldots,n_c$. The observables, including the
Hamiltonian, are real valued functions of $x$ and $k$, $A(x,k)$ (for
simplicity we disregard an explicit dependence on time in the observables).
Their dynamical evolution is described by the Poisson bracket with the
Hamiltonian
\begin{equation}
\frac{d}{dt}A = \{A,H\}
:=\sum_i
\left(
\frac{\partial A}{\partial x_i}\frac{\partial H}{\partial k_i}
-
\frac{\partial A}{\partial k_i}\frac{\partial H}{\partial x_i}
\right)
.
\label{eq:2.1}
\end{equation}

The (classical) canonical formalism seems particularly convenient in a
discussion about classical-quantum mixing because it has a parallel in the
quantum treatment. Indeed, in the Heisenberg picture quantum observables are
Hermitian operators with dynamical evolution dictated by Heisenberg's equation
of motion
\begin{equation}
\frac{d}{dt} A = \frac{1}{i\hbar}[A,H]
:= \frac{1}{i\hbar}(A H-H A)
.
\end{equation}

The classical bracket $\{,\,\}$ and the quantum one $[\,,\,]/i\hbar$, share
mathematical properties which are essential for the corresponding
dynamics. First, they are {\em Lie brackets}, that is, they are linear,
antisymmetric and fulfilling the Jacobi identity. Lie groups act through Lie
algebras, so the Lie bracket property is required in order to implement
observables as generators of groups of transformations, including the
dynamical evolution. For instance, if the bracket were not antisymmetric,
conservation of energy, $dH/dt=0$, would not be guaranteed
\cite{Diosi:1995fe}. Likewise, if the angular momentum is to generate
rotations of the system, the bracket has to carry a representation of the
algebra of SO(3), and so this bracket has to enjoy the Jacobi property.  Also,
this property ensures that a relation like $C=(A,B)$, where $ (\,,\,)$ denotes
the {\em dynamical bracket}, is preserved under dynamical evolution or other
transformations.

Throughout we will consider dynamics of ``universal'' type, rather than of
restricted type, so we really need the bracket between any two observables to
be defined (and to be itself an observable) since any observable can be
regarded as a possible Hamiltonian, or any observable can be added to the
Hamiltonian as a perturbation.

Another conspicuous property of the dynamical brackets is that they are a
{\em derivation}, i.e., they fulfill Leibniz's rule:
\begin{equation}
(A,BC)= (A,B)C+B(A,C)
.
\end{equation}
This property ensures that a relation like $C=AB$ is preserved under dynamical
evolution. For instance, if ${\bm p}$ is the momentum operator, the kinetic
energy ${\bm p}^2/2m$ evolves as ${\bm p}^2(t)/2m$. This avoids the odd
scenario in which the expression of the kinetic energy would be different at
different times, and similarly for any other observable without intrinsic time
dependence.

A further property refers to the structure of systems composed of different
sectors, i.e., different independent sets of degrees of freedom. For instance,
two different particles, or spin and position of a single particle. In this
case, the observables of the full system have the structure of {\em tensor
  product} over the various sectors. This is true in classical and in quantum
mechanics. An immediate consequence of the tensor product construction is that
observables of two different quantum sectors commute, and furthermore the
product of two such observables is also an observable (the product of two
Hermitian commuting operators being automatically Hermitian). The bracket of
two observables of the same sector remains in that same sector, and moreover,
the bracket of observables in two {\em different} sectors vanishes. This
property is important. It ensures that the two different sectors evolve
independently unless an interaction term is present in the
Hamiltonian. Indeed, if the Hamiltonian takes the form $H=H_1+H_2$, with $H_1$
and $H_2$ acting in the two different sectors, and the observable $A_1$
belongs to the first sector, its evolution will not depend on $H_2$.

The fact that the two canonical structures, classical and quantal, have common
properties is of course no accident. As is well known, using, e.g. the Wigner
representation \cite{Wigner:1932eb,Moyal:1949sk,Carruthers:1982fa}, the
Poisson bracket can be obtained as an $\hbar\to 0$ limit of the
commutator. The abovementioned properties are preserved by the limiting
procedure as they do not explicitly depend on $\hbar$.

A rather natural approach suggests itself to describe systems having
simultaneously quantum and classical degrees of freedom, namely, to start with
a quantum-quantum system and somehow take the classical limit in just one of
the two sectors.  (For instance, one could start with operators defined in the
tensor product Hilbert space of the two quantum sectors, apply a Wigner
transformation in just one of the spaces, and take the classical limit there.)
In that description observables would be Hermitian operators in the Hilbert
space of the quantum sector and also functions of the phase space variables of
the classical sector. 

Dynamical brackets have been proposed for the hybrid quantum-classical systems
in this approach, most notably by Aleksandrov and by Boucher and Traschen
\cite{Aleksandrov:1981,Boucher:1988ua}:
\begin{equation}
(A,B)= \frac{1}{i\hbar}[A,B]  + \frac{1}{2}\{A,B\} - \frac{1}{2}\{B,A\}
.
\end{equation}
(Here the Poisson bracket is applied as in \eq{2.1} to the operators $A$ and
$B$ which in general do not commute.)  This bracket has some good properties
(certainly better than the bracket proposed in \cite{Anderson:1995tn}, see
\cite{Diosi:1995fe,Jones:1996,Salcedo:1996jr}) but it is not a derivation and does not fulfill the Jacobi
identity. In addition, it does not preserve positivity of the density matrix
\cite{Boucher:1988ua}.

In fact no bracket can provide a dynamics with all the nice properties common
to the purely quantum or purely classical cases. (We refer to that hypothetic
dynamics as perfect hybridization.) This is shown in
\cite{Salcedo:1996jr}. The key point is that, although the properties hold for
any value of $\hbar$, they require $\hbar$ to take the same value in all
sectors \cite{Salcedo:1996jr,Caro:1998us}. Elaborated proofs based on this
idea can be found in \cite{Sahoo:2004,Dass:2009qu}. Here we present a simple
proof. Let $A_1$, $B_1$ be two observables in one sector and $A_2$, $B_2$ in
another sector.  Let us assume that the bracket $(\,,\,)$ in the total space
enjoys all the abovementioned properties, and in particular, in each sector
they are quantum brackets with two different Planck constants
\begin{equation}
 (A_1,B_1) =  \frac{1}{i\hbar_1}[A_1,B_1]
,
\qquad
 (A_2,B_2) =  \frac{1}{i\hbar_2}[A_2,B_2]
.
\label{eq:2.5}
\end{equation}

By assumption we can form the new observables $A_1 A_2$ and
$B_1 B_2$. Then, applying Leibniz's rule twice,
\begin{eqnarray}
(A_1A_2,B_1B_2) &=& 
(A_1A_2,B_1)B_2+B_1(A_1A_2,B_2)
\nonumber\\
&=&
 A_1(A_2,B_1)B_2 + (A_1,B_1)A_2B_2
+B_1A_1(A_2,B_2)+B_1(A_1,B_2)A_2
\nonumber\\
&=&
 (A_1,B_1)A_2B_2 + B_1A_1(A_2,B_2)
.
\end{eqnarray}
In the last equality it has been used that the bracket vanishes for different
sectors. On the other hand, due to antisymmetry of the bracket, the expression
is antisymmetric under exchange of labels $A$ and $B$:
\begin{equation}
(A_1A_2,B_1B_2) = - (B_1B_2,A_1A_2)
,
\end{equation}
therefore
\begin{eqnarray}
 (A_1,B_1)A_2B_2 + B_1A_1(A_2,B_2)
&=&
 -(B_1,A_1)B_2A_2 - A_1B_1(B_2,A_2)
\nonumber\\ 
&=&
(A_1,B_1)B_2A_2 + A_1B_1(A_2,B_2)
.
\end{eqnarray}
This implies the compact relation
\begin{equation}
 (A_1,B_1)[A_2,B_2] = [A_1,B_1](A_2,B_2)
.
\end{equation}
This relation for generic operators, combined with \eq{2.5}, leaves only the
possibility
\begin{equation}
\hbar_1=\hbar_2
.
\end{equation}
Hence there is no quantum-classical mixing (which would require
$\hbar_1=\hbar$, $\hbar_2=0$) with all the nice properties shared by the
purely classical or purely quantum cases. No perfect quantum-classical
mixing. Note that the Jacobi identity has not been used. Also Leibniz rule has
not been applied it is full power. We have only assumed that $A_1A_2$ evolves
into $A_1(t)A_2(t)$, i.e., only for the product of two observables in
different sectors.

Quantum-classical hybrids can be considered at two levels, a practical one and
a fundamental one. If quantum-classical hybrid systems are regarded as an
approximation to a full quantum system, the previous no-go theorem just shows
that such approximation will always meet some intrinsic limitations. This is
not particularly surprising and it does not prevent this kind of
approximations from being useful ones. On the other hand, if the aim is to
describe hypothetical quantum-classical hybrids truly existing in nature, the
no-go result implies that such hybrid systems will have {\em emergent}
features, not present in any of the two sectors separately. This is because a
hybrid with just the standard features has been shown not to be consistent. In
this scenario there are at least two alternatives. First, that quantum and
classical mechanics are just limit cases of a larger theory
\cite{Ghirardi:1986}, and in this case the emergent features were already
present from the beginning. Or second, whenever the two sectors, classical and
quantal, are not coupled by any interaction term in the Hamiltonian, they
behave precisely as expected from standard classical mechanics and from
standard quantum mechanics, being only their coupling what would yield new
emergent properties. To be practical, we will adopt the latter possibility as
our working assumption. Let us remark that the assumption refers not only to
the case of classical and quantum sectors which are never coupled, but also to
the cases in which the coupling acts occasionally. In support of this
assumption is the empirical fact that quantum mechanics is verified to work
very accurately for systems for which the previous history is not known (and so
they may include a previous interaction with hypothetic classical sectors).
Also, assuming that a quantum measurement requires a truly classical
apparatus, the assumption is supported by the fact that quantum mechanics
works accurately also after measurements have taken place.\footnote{Although
  this goes beyond the present discussion, the fact is that this author does
  not sympathize with the ``emergent scenario''.  In our view quantum
  mechanics would be the correct description of nature, not only a the atomic
  and subatomic levels but also at the macroscopic one, and classical
  mechanics would remain as just a limit case, a suitable approximation in
  many situations. That would most economically account for the fact that
  quantum and classical mechanics share a common structure, and moreover the
  same accuracy of quantum mechanics, overwhelmingly displayed in the micro
  world, would account for the accuracy of the classical description displayed
  in, e.g., celestial mechanics.}

\section{Statistical consistency and mean-field scheme}
\label{sec:3}

In this section we assume a system with truly quantum and truly classical
sectors, as described by their corresponding standard dynamics when they do
not interact. We show that {\em non linear} hybrid dynamics are in conflict
with quantum mechanics, as we understand it.\footnote{Obviously from the
  beginning there has been much debate about interpretation and other details
  of quantum mechanics. Here we refer to quantum mechanics as found in
  textbooks, e.g. \cite{Galindo:1991bk}.}

The simplest and most intuitive description of the quantum-classical mixing
follows from the well known {\em mean-field} dynamics. In this dynamics the
classical sector and the quantum sector remain (or can remain) always in pure
states. That is, at any time, and with or without interaction switched on, the
position and momentum of the classical particles are well defined, and the
quantum state is described by a wavefunction rather than a density matrix. The
dynamics is as follows
\begin{equation}
\frac{dx_i}{dt} = \frac{\partial }{\partial k_i}\langle H(x,k)\rangle_\psi
,
\quad
\frac{dk_i}{dt} = -\frac{\partial }{\partial x_i}\langle H(x,k)\rangle_\psi
,
\quad
i\hbar\frac{d}{d t}|\psi\rangle = H(x,k) |\psi\rangle
.
\label{eq:3.1}
\end{equation}
The Hamiltonian of the system is a function defined on the classical phase
space that takes values on operators of the Hilbert space of the quantum
system. Such dynamics contains back reaction of the quantum sector on the
classical sector, but misses the ``quantum back reaction'', that is the effect
of quantum fluctuation around the expectation value, that presumably should
also be present \cite{Wald:1994bk}.

A nice reformulation of the mean-field approach has recently been presented in
\cite{Elze:2011hi}.\footnote{Here we are representing our own point of
  view. It does not coincide with the point of view forwarded in
  \cite{Elze:2011hi}.} This is based on the well known observation that the
wave function can be regarded as a classical field, and the Schr\"odinger
equation can be regarded as the corresponding classical field equation of
motion.  Quantum observables can be represented by their expectation value,
$\mathcal{A}(\psi)= \langle A\rangle_\psi$, so that the commutator is
represented by the Poisson bracket, with $\psi_q$ and $i\hbar\psi_q^*$ as the
canonical conjugate variables. Here $\psi_q=\langle q|\psi\rangle$ and
$|q\rangle$ is any orthonormal basis of the Hilbert space of the quantum
sector. In the hybrid case these variables are augmented with the phase space
variables of the classical sector, and $\mathcal{A}(\psi,x,k)= \langle
A(x,k)\rangle_\psi$,
\begin{equation}
\{\mathcal{A},\mathcal{B}\}
=
\frac{1}{i\hbar}\sum_q
\left(
\frac{\partial \mathcal{A}}{\partial \psi_q}
\frac{\partial \mathcal{B}}{\partial \psi^*_q}
-
\frac{\partial \mathcal{A}}{\partial \psi^*_q}
\frac{\partial \mathcal{B}}{\partial \psi_q}
\right)
+
\sum_i
\left(
\frac{\partial \mathcal{A}}{\partial x_i}
\frac{\partial \mathcal{B}}{\partial k_i}
-
\frac{\partial \mathcal{A}}{\partial k_i}
\frac{\partial \mathcal{B}}{\partial x_i}
\right)
.
\end{equation}

The hybrid dynamics given by
\begin{equation}
\dot{X} = \{X,\langle H(x,k)\rangle_\psi\}
,
\quad
X=\psi_q,\psi^*_q,x_i,k_i
,
\end{equation}
is easily shown to be equivalent to that in \eq{3.1}.

In this formulation there is a Lie bracket which is also a derivation, so this
approach would seem to bypass the no-go theorem. The caveat is that the
dynamical bracket of two observables should be itself an observable, and this
is not the case here. In the scheme of \cite{Elze:2011hi} observables are
expectation values of operators $A(x,k)$, and so bilinear in $\psi$ and
$\psi^*$. This property is not preserved by the classical part of the bracket
(which in general will be quadratic in $\psi$ and quadratic in $\psi^*$). This
implies that the time derivative of an observable, $\dot{\mathcal{A}}$, is not
an observable, i.e., of the form $\mathcal{B}= \langle B(x,k)\rangle_\psi$,
for some operator valued $B(x,k)$.

Now we come to the main argument of this work. We introduce a new consistency
condition to be added to other considered up to now in the literature. In
quantum mechanics, as commonly understood, the state of a system is described,
in the most general case, by a density matrix
\cite{vonNeumann:1927,Galindo:1991bk} (interpreted in the usual sense of
``proper mixtures'' \cite{dEspagnat:1995}).  This represents a statistical
mixture of pure states, pure states themselves being a particular case.  The
key observation is the well known fact that, in general, density matrices can
be realized in many different ways as mixtures of pure states.  A simple
example is that of an unpolarized electron beam. Such state can be attributed
to an equiprobable mixture of up and down spins, but the same mixture is
obtained regardless of the quantization axis chosen.  For another example, let
\begin{equation}
\hrho = 
\sum_\alpha p_\alpha|\psi_\alpha\rangle \langle \psi_\alpha|
,
\
\quad
p_\alpha \ge 0
,
\quad
\sum_\alpha p_\alpha = 1
,
\label{eq:3.4}
\end{equation}
where the $|\psi_\alpha\rangle$ are normalized but not orthogonal. Because
$\hrho$ is Hermitian and positive it can be diagonalized into orthonormal
states with positive weights
\begin{equation}
\hrho = 
\sum_\nu w_\nu|\phi_\nu\rangle \langle \phi_\nu|
,
\quad
w_\nu \ge 0
,
\quad
\sum_\nu w_\nu = 1
,
\quad
\langle \phi_\nu|\phi_{\nu^\prime}\rangle = \delta_{\nu\nu^\prime}
.
\end{equation}
The new states (eigenstates of $\hrho$) are linear combination of the old
ones, but different from them (unless all the $|\psi_\alpha\rangle$ are the
same state).

In general, we can consider that in \eq{3.4} the label $\alpha$ runs through the
set of {\em all} pure states $|\psi_\alpha\rangle$ (normalized vectors, and
modulo a phase) each pure state with some weight $p_\alpha$.  Note that we
mean all states, not just a linear basis of states.  This is an infinite
number even for a qubit. A configuration $\{p_\alpha\}$ will produce a density
matrix, but the number of possible different density matrices is much smaller
than that of configurations. All the configurations $\{p_\alpha\}$ yielding
the same density matrix represent precisely the same quantum state. In quantum
mechanics there is no way to distinguish between two mixtures producing the
same density matrix. Not only the expectation value of every observable will
be the same, $\tr(\hrho A)$, but also the results of any measurement will
be identical, as also the probabilities can be written using the density
matrix only, $P(A=a)=\langle
a|\hrho|a\rangle$~\cite{Galindo:1991bk}. This means that, in quantum
mechanics, the precise decomposition of a density matrix into pure states has
no physical meaning.

In the classical theory there is the probability density function on phase
space $\rho(x,k)$ and in this case the decomposition into pure states
$\delta(x-a)\delta(k-b)$ is unique,
\begin{equation}
\rho(x,k) = \int d^n a \, d^n b \, \rho(a,b)\delta(x-a)\delta(k-b)
.
\end{equation}

A classical-quantum hybrid scheme like the mean-field one, does not directly
dictate an evolution for statistical mixtures of classical or quantal pure
states. However, nothing prevents us from applying the hybrid scheme for pure
states and take the statistical mixing at any time. The basis for this
procedure follows from the meaning of the statistical mixture and from
standard probability theory. It does not rely on quantum mechanics.  This
implies a stringent consistency condition on any hybrid scheme. We may not know
how things work in a hybrid system when they interact, but we know that in the
absence of interaction each sector behaves in the standard way. Therefore, let
the interaction be switched off for $t<t_0$, and let the state at $t_0$ be
$(x,k)=(x_0,k_0)$ in the classical sector and $|\psi_\alpha\rangle$ in the
quantum sector. The interaction is connected for $t>t_0$ and the evolution
depends on the hybrid scheme adopted. In fact, we can consider all such
evolutions for all possible initial $|\psi_\alpha\rangle$ (but the same
$(x_0,k_0)$). Let $\mathcal{A}_\alpha(t)$ be the expectation value of any
hybrid observable $A$ for each $i$ at time $t$.  Whenever two mixtures
$\{p_\alpha\}$ and $\{p^\prime_\alpha\}$ produce the same density matrix at
$t=t_0$
\begin{equation}
\hrho = 
\sum_\alpha p_\alpha|\psi_\alpha\rangle \langle \psi_\alpha|
=
\sum_\alpha p^\prime_\alpha|\psi_\alpha\rangle \langle \psi_\alpha|
,
\end{equation}
we should demand that
\begin{equation}
\sum_\alpha p_\alpha \mathcal{A}_\alpha(t) 
= \sum_\alpha p^\prime_\alpha \mathcal{A}_\alpha(t)
\qquad \mbox{for $t \ge t_0$ and for any observable ~$A$}
.
\label{eq:3.8}
\end{equation}
The reason is that for $t\le t_0$ the two states described by $\{p_\alpha\}$
and $\{p^\prime_\alpha\}$ are identical, if quantum mechanics is correct and
complete for the isolated quantum sector.  Therefore the evolution at later
times of the two states should also be identical, regardless of the nature of
the hybrid dynamics.

A hybrid scheme violating the condition contained in \eq{3.8} would
automatically provide an experimental way to discriminate between two mixtures
which, according to quantum mechanics, are indistinguishable, and presumably
would provide, for any mixture, its ``true'' decomposition into pure states.
Such ``true'' pure states would be a type of hidden variables in quantum
mechanics, as quantum mechanics is blind to them. This possibility should be
rejected: if such hybrid scheme is applied to describe the measurement process
in quantum mechanics, the measurement apparatus being classical, the fact that
the scheme distinguishes mixtures with the same density matrix would be in
contradiction with what is known about expectation values and measurements in
quantum mechanics. So a hybrid system violating \eq{3.8} is either
inconsistent or of limited applicability. For want of a better name, we refer
to the requirement just introduced as {\em statistical consistency}.

The mean-field scheme violates this condition. To see this, consider a
collection of alternative pure states $|\psi_\alpha\rangle$ at $t_0$ in the
quantum sector with a common state $(x_0,k_0)$ in the classical sector, and
let each hybrid state have its evolution. In the mean-field dynamics, the
evolution of the expectation of an observable takes the form
\begin{equation}
\frac{d}{dt}\langle A \rangle_\alpha
=
\frac{1}{i\hbar}
\langle [A , H] \rangle_\alpha
+
\{\langle A \rangle_\alpha, \langle H \rangle_\alpha \}
.
\label{eq:3.9}
\end{equation}
Multiplying by $p_\alpha$, summing over $\alpha$ and taking $t=t_0$, we can
see that the last term can not be expressed in terms of $\sum_\alpha
p_\alpha|\psi_\alpha(t)\rangle \langle\psi_\alpha(t)|$, due to the lack of
linearity in the dynamics. Hence statistical consistency is lost.

For instance, for a classical particle with quantum spin and Hamiltonian
$H=\frac{\lambda}{2}\vx^2\vk\!\cdot\!\vsigma$, let the state at $t_0$ be
$(\vx_0,\vk_0)$, and unpolarized spin.  To implement this, let us choose an
arbitrary axis $\hat{\vn}$ and let the pure spin states be
$|\!\uparrow\rangle$ and $|\!\downarrow\rangle$, each with probability one
half. At $t_0$ one finds $\frac{d}{dt}\langle\vk\!\cdot\!\vsigma\rangle=
-\lambda\hat{\vn}\!\cdot\!\vx_0\hat{\vn}\!\cdot\!\vk_0$.  The dependence on
$\hat{\vn}$ breaks statistical consistency.\footnote{This example suggests to
  restore statistical consistency in the mean-field approach by making a
  suitable average over all possible decompositions of a given density matrix
  into pure states, $\hrho=\sum_\alpha p_\alpha|\psi_\alpha\rangle
  \langle\psi_\alpha|$. In the example above, a rotational average over
  $\hat{\vn}$ seems appropriate. It is not clear to us whether, in the general
  case, there is a natural density probability function defined on the set of
  choices $p_\alpha$, and to what extent this procedure would be an
  improvement regarding consistency.}

The mean-field evolution in \eq{3.9} can be also written as
\begin{equation}
\frac{d}{dt}\tr (\hat\rho A )
=
\frac{1}{i\hbar}
\tr (\hat\rho  [A , H] )
+
\{\tr (\hat\rho A ), \tr (\hat\rho H ) \}
,
\label{eq:3.10}
\end{equation}
where $\hat\rho=|\psi_\alpha(t)\rangle \langle\psi_\alpha(t)|$ (no sum over
$\alpha$). The equation written in this form suggests to propose this very
dynamics but now inserting in $\hrho$ a general density matrix. In this case
$\hrho$ would evolve according to
\begin{equation}
\frac{dx_i}{dt} = \frac{\partial }{\partial k_i}\tr(\hrho H )
,
\quad
\frac{dk_i}{dt} = -\frac{\partial }{\partial x_i}\tr(\hrho H )
,
\quad
i\hbar\frac{d}{d t}\hrho = [H,\hrho]
.
\end{equation}
Unfortunately such evolution is not consistent with the meaning of statistical
mixture. That meaning implies that, given two alternative situations $\hrho_1$
and $\hrho_2$ with probabilities $p_1$ and $p_2$, the mixture $\hrho = p_1
\hrho_1 + p_2 \hrho_2$ at $t=t_0$ should remain so at any other time.  Each
alternative represents a possible different history and, by definition of
expectation value, one should have a weighted average of the two histories,
$\langle A\rangle_\rho= p_1\langle A\rangle_{\rho_1} + p_2\langle
A\rangle_{\rho_2}$ at any time. So in practice linearity in $\hrho$ is
required and this constraint is not fulfilled by \eq{3.10}.\footnote{Strictly
  speaking, linearity means preservation of the relation $\hrho = \lambda_1
  \hrho_1 + \lambda_2 \hrho_2$, for any real weights $\lambda_{1,2}$, while we
  only need non negative weights $p_{1,2}$.}

Hybrid schemes like those considered in Section \ref{sec:2}, where observables
are operator valued functions in the classical phase space, meet the
requirement of statistical consistency. They can be formulated using the
combined density matrix of quantum and classical sectors, $\hrho(x,k)$, with
the following linear evolution
\begin{equation}
\frac{d}{dt}\hrho(x,k) = (\hat{H}(x,k),\hrho(x,k))
.
\end{equation}
Therefore the issue of the manifold decomposition of the density matrix into
pure states is never raised. (But these dynamics are subject to the no-go
theorem of Section \ref{sec:2}.) On the other hand, hybrid schemes which do
not preserve the linearity of quantum mechanics are likely to have trouble
with the requirement of statistical consistency. Conflicts with the principle
of locality have been also observed \cite{Gisin:1990}.

\section{Statistical ensemble approach}
\label{sec:4}

\subsection{The scheme}

In the statistical ensemble approach of \cite{Hall:2005ax,Hall:2008}, the
basic state of a quantum-classical hybrid system is described by two real
functions, $P(x,q)$ and $S(x,q)$, defined on configuration space, $x$ being
the classical coordinates and $q$ the quantum ones. $P(x,q)$ represents the
probability density function of the state $(x,q)$ and so it is non negative
and normalized.\footnote{Following \cite{Hall:2005ax,Hall:2008}, we use $x$,
  $q$, $\int dx\,dq$, etc, although $x$ and $q$ represent sets of several
  coordinates.}

When the quantum sector is missing, the pair $P(x)$ and $S(x)$ represents a
particular type of mixed state, namely, that with phase space probability
density function
\begin{equation}
\rho(x,k) = P(x)\delta(k-\nabla S(x))
.
\label{eq:4.1}
\end{equation}
This form is preserved by the purely classical dynamics. Using Hamilton's
equations to evolve $x$ and $k$, one finds the evolution of $P$ and $S$.
For a classical particle with mass $M$ in presence of a potential $V(x,t)$,
\begin{equation}
\frac{\partial P}{\partial t}
=
-\frac{1}{M}\nabla(P\nabla S)
,
\qquad
\frac{\partial S}{\partial t}
=
-\frac{1}{2M}(\nabla S)^2-V
.
\end{equation}
The first relation is the continuity equation and the second one is the
Hamilton-Jacobi equation. They can be derived from a canonical bracket. For
two functionals of $P$ and $S$,  $\mathcal{A}$ and $\mathcal{B}$ 
\begin{equation}
\{\mathcal{A},\mathcal{B} \}
=
\int dx \left(
\frac{\delta\mathcal{A}}{\delta P(x)}
\frac{\delta\mathcal{B}}{\delta S(x)}
-
 \frac{\delta\mathcal{A}}{\delta S(x)}
\frac{\delta\mathcal{B}}{\delta P(x)}
\right)
.
\end{equation}
Hence $d\mathcal{A}/d t = \{\mathcal{A},\mathcal{H}\}$, with Hamiltonian
\begin{equation}
\mathcal{H} =  \int dx P \left( \frac{1}{2M}(\nabla S)^2 + V \right)
.
\end{equation} 

Likewise, when the classical sector is missing, the quantum state is a pure
state with wave function
\begin{equation}
\psi(q)= P(q)^{1/2}e^{iS(q)/\hbar}
.
\end{equation}
For a quantum particle with mass $m$ in a potential $V(q,t)$, the
Schr\"odinger equation evolves the pair $P(q)$ and $S(q)$ as
\begin{equation}
\frac{\partial P}{\partial t}
=
-\frac{1}{m}\nabla(P\nabla S)
,
\qquad
\frac{\partial S}{\partial t}
=
-\frac{1}{2m}(\nabla S)^2
+
\frac{\hbar^2}{2m}\frac{\nabla^2 P^{1/2}}{P^{1/2}}
-V
.
\label{eq:4.6}
\end{equation}
This evolution derives from a bracket similar to the classical one (with $q$
instead of $x$), this time with Hamiltonian
\begin{equation}
\mathcal{H} =  \int dq\, P \left( \frac{1}{2m}(\nabla S)^2 
+
\frac{\hbar^2}{8m}(\nabla \log P)^2
+ V 
\right)
.
\end{equation}
So in this approach the quantum description differs from the classical one
just by the term with explicit $\hbar$ in $\mathcal{H}$.

The guiding principle to introduce hybrid quantum-classical systems is that
new degrees of freedom are to be added exactly in the same way as it is done
in the purely classical or purely quantum cases, namely, by adding new
coordinates in the functions $P$ and $S$. The hybrid case is then described
by $P(x,q)$ and $S(x,q)$, and the dynamical bracket is 
\begin{equation}
\{\mathcal{A},\mathcal{B} \}
=
\int dx \, dq \left(
\frac{\delta\mathcal{A}}{\delta P(x,q)}
\frac{\delta\mathcal{B}}{\delta S(x,q)}
-
 \frac{\delta\mathcal{A}}{\delta S(x,q)}
\frac{\delta\mathcal{B}}{\delta P(x,q)}
\right)
.
\end{equation}

For classical and quantum particles interacting through a potential $V(x,q,t)$
\begin{equation}
\mathcal{H} =  \int dx\, dq \, P \left( 
\frac{1}{2M}(\nabla_x S)^2 
+
\frac{1}{2m}(\nabla_q S)^2 
+
\frac{\hbar^2}{8m}(\nabla_q \log P)^2
+ V 
\right)
,
\end{equation}
and this produces the hybrid evolution equations
\begin{equation}
\frac{\partial P}{\partial t}
=
-\frac{1}{M}\nabla_x(P\nabla_x S)
-\frac{1}{m}\nabla_q(P\nabla_q S)
,
\qquad
\frac{\partial S}{\partial t}
=
-\frac{1}{2M}(\nabla_x S)^2
-\frac{1}{2m}(\nabla_q S)^2
+
\frac{\hbar^2}{2m}\frac{\nabla_q^2 P^{1/2}}{P^{1/2}}
-V
.
\label{eq:4.10a}
\end{equation}

The scheme extends straightforwardly to more general quantum systems (e.g.,
with spin degrees of freedom).

In this approach observables are represented by their expectation value, as
real valued functionals of $P$ and $S$. (The Hamiltonians above follow this
rule.)  So a classical observable $f(x,k)$ is represented by the functional
\begin{equation}
\mathcal{F} = \int dx\, dq\, P f(x,\nabla_x S)
,
\label{eq:4.10}
\end{equation}
whereas a quantum observable $\hat{A}$ is represented by the functional
\begin{equation}
\mathcal{A} = \int dx\, \langle\psi(x)|\hat{A} |\psi(x)\rangle
,
\qquad
\psi(x,q) = \langle q|\psi(x)\rangle = P(x,q)^{1/2} e^{i S(x,q)/\hbar}
.
\label{eq:4.12}
\end{equation}

The approach passes a number of tests listed in
\cite{Hall:2005ax,Hall:2008,Reginatto:2009}. However, it has some limitations
too.  There is a problem in the definition of which functionals
$\mathcal{A}[P,S]$ are acceptable as observables. The statistical
interpretation of $P$ requires the observables to be homogeneous functionals
of degree one in $P$.  Some further constraints are noted in
\cite{Hall:2005ax,Hall:2008} to ensure positivity of $P$ during the dynamical
evolution and global phase invariance of the wave function (this is required
to implement gauge invariance and so a consistent electromagnetic
coupling).\footnote{For classical and quantum particles with charge $Q_1$ and
  $Q_2$, minimal coupling is achieved by the replacements $\nabla_x S\to
  \nabla_x S - Q_1 A(x,t)$, $\nabla_q S\to \nabla_q S - Q_2 A(q,t)$, $V\to
  V+Q_1\phi(x,t)+Q_2\phi(q,t)$ in $\mathcal{H}$. The dynamics is invariant
  under the gauge transformations: $P(x,q,t)\to P(x,q,t)$, $S(x,q,t)\to
  S(x,q,t)+Q_1 \Lambda(x,t)+Q_2 \Lambda(q,t)$, $A\to A+\nabla \Lambda$,
  $\phi\to \phi-\partial_t \Lambda$. Observe that the electromagnetic field is
  not dynamical here.}  But these conditions still leave an enormous set of
functionals. To see that restrictions are needed, consider a purely classical
system. There, a term
\begin{equation}
\mathcal{H}_1 =  \int dx\, P(x)
\frac{\hbar^2}{8M}(\nabla \log P(x))^2
\end{equation}
added to the Hamiltonian passes all the noted conditions, but still it is not
acceptable, because in classical mechanics we know which functionals are true
observables (to wit, those in \eq{4.10} upon removal of $q$). Note that the
``interaction'' $\mathcal{H}_1$ would turn the classical particle into a
quantum one. Likewise a quantum particle could be turn into a classical one by
switching on a suitable ``interaction'' term.

The previous argument suggests that also in the hybrid case most functionals
are completely unrelated to observables. In the purely classical or purely
quantum cases, the growth in the number of observables when new degrees of
freedom are added is limited. As noted above, new observables in the full
system are obtained by tensor product of the observables in the
subsectors. However the product of a classical observable with a quantum
observable is not automatically defined in the ensemble approach to hybrid
systems: the blocks $f(x,\nabla_xS)$ and $\hat{A}$ do not commute in general
($\hat{A}$ acts on $q$ and $S$ contains $q$). Of course, one could introduce
some symmetrization prescription (plus possible $O(\hbar)$ terms), thus
defining a set of hybrid observables.\footnote{Note that once the product of a
  classical observable, $A_1$, with a quantum observable, $A_2$, is defined,
  the product of general observables is straightforwardly defined by
  $(A_1A_2)(B_1B_2)= (A_1B_1)(A_2B_2)$.} However this is not sufficient
because it should be verified that such set is closed upon application of the
dynamical bracket, and this condition is far from trivial.

Another problem of the ensemble approach is that the bracket of a generic
quantum observable with a generic classical observable is not
zero~\cite{Hall:2008}.  This is a serious trouble because, as noted above, it
implies that the classical Hamiltonian induces an evolution in the quantum
sector and vice versa, even when no interaction is taking place between both
sectors.  Such bracket vanishes only for particular observables or for
particular configurations~\cite{Hall:2008}. An instance of such special
configuration is the separable one:
\begin{equation}
P(x,q,t)=P(x,t)P(q,t)
,
\qquad
S(x,q,t) = S(x,t)+S(q,t)
.
\end{equation}
(Of course, the various functions $P$ and the various functions $S$ are
different, for short we let their arguments to distinguish them.) The
separable case represents sectors which never interact. It is easy to verify
that, in the absence of interaction between sectors, the separable form is
preserved by the evolution. In the separable case the classical sector does
not act on the quantum one and vice versa. However, this is not sufficient for
a consistent dynamics. If the interaction between sectors is switched on
during a certain time interval and then set to zero, the configuration will no
longer be separable, the two sectors are {\em entangled}. Then one would find
that, even though the two sectors are no longer connected, what happens to one
sector affects to the other.

A concrete observable affected by such ghost coupling induced by ``hybrid
entanglement'' is the kinetic energy. For instance, according to the ensemble
approach, if a neutral {\em free} classical particle is ``entangled'' with a
charged quantum one, an electromagnetic wave acting upon the quantum particle
would induce a variation with the same frequency in the energy of the
classical particle. By all accounts, the kinetic energy of a classical
particle (if such thing exists in nature) would be a {\em bona fide}
observable quantity and so the ghost coupling is odd.  The roles of quantum
and classical can be exchanged with a similar conclusion.  It should be noted
that the entanglement does not produce such effect in the quantum-quantum case
or in the classical-classical case. In those cases the bracket of different
sectors vanishes, as one would expect.

Instead of using the non vanishing of the bracket, the same effect can be seen
from the evolution of the marginal probability distribution of $(x,k)$
\begin{equation}
\rho(x,k) = \int dq\,P(x,q)\,\delta(k-\nabla_xS(x,q))
.
\end{equation}
In the separable case this reduces to $P(x)\,\delta(k-\nabla_xS(x))$, so what
happens to the quantum sector has no effect on the classical sector.  But in
the general entangled case it would seem that the evolution of $q$ will
produce some effect even if there is no interaction. In fact, this is not so
straightforward as it would seem. Note that the same naive argument would
apply in the classical-classical case. However, in that case, using the
evolutions of $P(x,q)$ and $S(x,q)$ (with $\hbar\to 0$) it can be shown that
the net effect vanishes because it comes in the form of a total derivative
with respect to $q$. In the quantum-classical case a similar cancellation
occurs for the marginal distribution of $x$ (hence for observables depending
only on $x$ and not on $k$) and also for the special case of $\langle
k\rangle$ \cite{Hall:2008}, but not for the full $\rho(x,k)$, due to the extra
term $\hbar^2/(2m)\nabla_q^2P ^{1/2}/P ^{1/2}$ in the evolution of $S(x,q)$.

There is a related difficulty with conservation of angular momentum in the
presence of spin. The simplest setting to show this is a classical particle
$(\vx,\vk)$ with quantum spin $1/2$.\footnote{Everything can be repeated, with
  the same conclusions, for a classical particle interacting with a quantum
  particle having spin $1/2$. In this case we would have $\psi_a(x,q)$. The
  case considered in the text would follow by assuming that the motion of the
  quantum particle can be neglected.} The wave function of the hybrid system
(analogous to $\psi(x,q)$ in \eq{4.12})
is
\begin{equation}
\psi_\alphaa(\vx) = P_\alphaa^{1/2}(\vx)e^{i S_\alphaa(\vx)/\hbar}
.
\end{equation}
In the present case $q=\alphaa= \pm\frac{1}{2}$ is a discrete label, so we put
it as a subindex. $\hbar\alphaa$ is the projection of the spin along some
quantization axis. In order to define the angular momentum, we should specify
the action of the rotation group. The obvious way is to regard
$\psi_\alphaa(\vx)$ as a bispinor with respect to $\alphaa$ (since this is the
only choice when the classical sector is not present). Hence, for a
rotation $\vomega= \phi\hat{\vn}$,
\begin{equation}
\psi_\alphaa(\vx) \to (e^{-i\vomega \vsigma/2}\psi)_\alphaa(R^{-1}\vx)
.
\end{equation}
This procedure correctly defines a representation of the rotation group, and
the generator can be realized (using the dynamical bracket) by means of an
observable:
\begin{equation}
\vJ 
= \vL + \vS
=
\int d^3x \sum_{\alphaa,\betaa}
\psi_\alphaa^*(\vx)
\big(
-i\hbar \, \vx \times \vnabla
\delta_{\alphaa\betaa}
+
\frac{\hbar}{2}\vsigma_{\alphaa\betaa}
\big)
\psi_\betaa(\vx)
.
\end{equation}
Note that the orbital (classical) part can be written equivalently in the
form
\begin{equation}
\vL = 
\int d^3x \sum_\alphaa P_\alphaa(\vx) \, \vx \times \vnabla S_\alphaa(\vx)
.
\end{equation}
So $\vL$ and $\vS$ are the angular momenta observables to be expected for the
classical and for the quantum systems. In the present case, the symmetry group
guarantees that the bracket of $\vL$ with $\vS$ vanishes and that $\vJ$,
$\vL$, $\vS$ are all of them angular momenta, i.e., fulfill the commutation
relations of the so(3) algebra. The bracket of $\vJ$ with an observable
yields the effect of an infinitesimal rotation on the observable.

Let the hybrid particle be free. The Hamiltonian contains just the kinetic
energy of the classical sector,
\begin{equation}
\mathcal{H} =  \int d^3x \sum_\alphaa
 \frac{1}{2M}P_\alphaa(\vx)(\vnabla S_\alphaa(\vx))^2 
.
\label{eq:4.20}
\end{equation} 

Note that everything is fixed and there is no freedom to change anything.
Unfortunately, $\mathcal{H}$ is not at all an invariant functional under spin
rotations. $P_\alphaa(\vx)$ and $S_\alphaa(\vx)$ do not have good
transformation properties under rotations. In particular, they are not
bispinors.  This means that for each choice of quantization axis,
$\mathcal{H}$ represents a different functional.\footnote{Once again one could
  consider taking an average over all choices of axis, hence restoring the
  rotational invariance. However the resulting Hamiltonian would contain
  $\vsigma$ and so it would represent an interaction between the two sectors
  rather than a free classical particle.}  As can be shown in detail, the
bracket $\{\vJ,\mathcal{H}\}$ is not zero, so the Hamiltonian is not
rotationally invariant, and conversely, the angular momentum is not
conserved. ($\vL$ is conserved but $\vS$ is not.)  Of course, if $(\vx,\vk)$
and $\alphaa$ are not entangled, $P_\alphaa(\vx)=P(\vx)P_\alphaa$,
$S_\alphaa(\vx)=S(\vx)+S_\alphaa$, these problems do not arise. It is
instructive to note that the free {\em quantum-quantum} Hamiltonian,
\begin{equation}
\int d^3x \sum_\alphaa \frac{\hbar^2}{2M} |\vnabla\psi_\alphaa(\vx)|^2
=
\int d^3x \sum_\alphaa
 \left( 
\frac{1}{2M}P_\alphaa(\vx)(\vnabla S_\alphaa(\vx))^2  
+
\frac{\hbar^2}{8M}\frac{(\vnabla P_\alphaa(\vx))^2}{P_\alphaa(\vx)}
\right)
,
\end{equation}
is rotationally invariant but the two terms in the right-hand side are not
separately invariant.  (The first term, without $\hbar$, is the classical
Hamiltonian of \eq{4.20}.)

The problem seems to be ubiquitous for the implementation of internal
symmetries in the quantum sector, due to the non linear nature of the
approach. This can be seen from the modified Schr\"odinger equation obeyed by
the extended wave function $\psi_\alphaa(x,q)$ of the hybrid system, $\alphaa$
being a generic internal index,
\begin{equation}
i\hbar \frac{\partial}{\partial t}\psi_\alphaa
=
\left(
-\frac{\hbar^2}{2M}\nabla_x^2
-\frac{\hbar^2}{2m}\nabla_q^2
+\frac{\hbar^2}{2M}\frac{\nabla_x^2|\psi_\alphaa|}{|\psi_\alphaa|}
+V\right)
\psi_\alphaa
.
\label{eq:4.22}
\end{equation}
The effective potential $V_\alphaa(x,q)=
(\hbar^2/2M)\nabla_x^2|\psi_\alphaa|/|\psi_\alphaa|$, that turns $x$ into a
classical degree of freedom, tends to break any internal symmetry carried by
the index $\alphaa$. This is also an impediment to describe relativistic
particles with spin. The same problem pointed out for spin-$1/2$ particles
reappear if on tries to couple the electromagnetic field to classical charged
particles, if the photon is a quantum dynamical degree of freedom.

All the difficulties noted for the statistical ensemble approach have a common
root. They stem from the hybrid description
\begin{equation}
\psi(x,q) = P(x,q)^{1/2} e^{iS(x,q)/\hbar}
.
\end{equation}
Here $q$ represents the configuration label of a basis $|q\rangle$ of the
Hilbert space of the quantum system. The separable case describes two sectors
that never interact. This case is of limited interest, so we consider the
entangled case.  If $x$ and $q$ are entangled, a change of basis $|q\rangle$
affects in a non trivial way the other sector: it modifies the marginal
distribution of the classical sector and in particular its kinetic energy.

Formally, the pair $(P(x,q),S(x,q))$ carries the same information as
$\psi(x,q)$, but the former description is better suited for the purely
classical case and the second description is better suited for purely quantum
case (just consider how linear superposition of quantum states reflects on the
pair $(P,S)$). Using a common language for the quantum and classical sectors,
by means of any of these two descriptions for the classical-quantum case, does
not by itself solve the problem of quantum-classical hybridization. A
canonical transformation in the classical-classical case acts naturally on
$(P(x,q),S(x,q))$, but a unitary transformation in $q$ acts
awkwardly. Likewise, a classical canonical transformation on $x$ acts in an
unnatural way on $\psi(x,q)$ regarded as a wave function.

If $q$ does not have a classical limit, as happens for the spin or other
internal labels, there is no natural choice of basis $|q\rangle$ and each
choice produces different evolution even with no interaction present between
(entangled) sectors. However, even when it seems that there is a privileged
choice of basis, like $q$ labeling the position of a quantum particle, the
problem remains.  Indeed, the dynamical evolution is nothing else than a
continued canonical transformation, produced by the dynamical bracket. In the
quantum case this is a continued change of orthonormal basis. In the classical
case it is a continued change of canonical coordinates.  Hence, the root of
the difficulties is that the two types of canonical transformations are not
compatible because the two sectors carry two different values of $\hbar$.

\subsection{Statistical consistency of the scheme}

In this subsection we examine the statistical consistency of the ensemble
approach to hybrid systems. As discussed in Section \ref{sec:3} a non linear
scheme is unlikely to fulfill statistical consistency. This is the case of the
statistical ensemble approach, and in fact the situation is even worse than
that found for the mean-field approach. For the mean-field dynamics the
failure came from the {\em quantum} version of the statistical consistency
requirement, that is, different evolutions were obtained from different
decompositions of a single quantum density matrix. In the statistical ensemble
approach the failure takes place even for the {\em classical} version of the
requirement (as well as for the quantum one). This means the following. The
statistical ensemble approach is based on the evolution of classical ensembles
of the type $(P(x),S(x))$ in \eq{4.1}. This by no means represents the most
general classical ensemble, $\rho(x,k)$. In turn, a generic $\rho(x,k)$ can be
decomposed in many different ways as a combination of ensembles $(P,S)$,
\begin{equation}
\rho(x,k) = \sum_\alpha p_\alpha P_\alpha(x)\,\delta(k-\nabla S_\alpha(x))
.
\label{eq:4.23}
\end{equation}
This follows from the fact that $\alpha$ runs over the set of all possible
pairs $(P(x),S(x))$ and so the number of possible $p_\alpha$ is much larger
than that of possible probability density functions $\rho(x,k)$. Because only
$\rho(x,k)$ is meaningful, it should be demanded that different decompositions
produce the same evolution. This requirement is of course true in the purely
classical case ($\rho(x,k,t)$ fulfills the autonomous Liouville equation which
is consistent because is linear) but it fails to hold for the hybrid
evolution.\footnote{Pure classical states $(x_\alpha,k_\alpha)$ could be used
  to make a standard decomposition of $\rho(x,k)$, since they can be cast in
  the $(P,S)$ form, e.g. with $P_\alpha(x)=\delta(x-x_\alpha)$ and
  $S_\alpha(x)=k_\alpha x$. The problem (apart from $P_\alpha(x)$ being too
  singular for using it in the hybrid dynamics \eq{4.10a}) is that, for
  generic interactions, the form of $S_\alpha(x)$ is not preserved by the
  classical evolution, so one should be prepared to proof that the evolution
  does not depend on the concrete of choice of $S_\alpha(x)$. Besides, the
  lack of statistical consistency in the quantum sense remains.}

To show this, we take the non controversial case of a classical particle and a
quantum particle without internal degrees of freedom, interacting through a
potential $V(x,q,t)$. To test statistical consistency we can consider the
expectation value of observables of the form $\mathcal{A}(t)=\int
dx\,dq\,P(x,q,t)\,A(x,q,t)$. These are hybrid observables and should be
admissible since the potential belongs to this class. Of course, we can trade
all these observables by the probability density $P(x,q,t)$. So $P(x,q,t)$ is
itself an observable.

We assume a set of possible histories labeled by an index $\alpha$, each
history with probability $p_\alpha$. At $t=t_0$ all hybrid states are
separable with a common quantum state $\psi(q)$, and classical state described
by a pair $(P_\alpha(x),S_\alpha(x))$. For simplicity we assume a time
independent potential $V(x,q)$ for $t>t_0$.  The expectation value of the
observables of the type described above, taken over the set of histories,
depends on the probability density
\begin{equation}
P(x,q,t) = \sum_\alpha p_\alpha P_\alpha(x,q,t)
,
\label{eq:4.25}
\end{equation}
where $P_\alpha(x,q,t)$ is the probability density of the history $\alpha$, as
obtained by the hybrid evolution of the ensemble approach.

Statistical consistency requires that $P(x,q,t)$ should depend only on the
initial classical probability density function $\rho(x,k)$, and not on its
decomposition into histories, \eq{4.23}. We will call {\em statistical
  invariants} the quantities which are independent of the concrete
decomposition.  Hence, $\rho(x,k)$ as well as $\psi(q)$ and $V(x,q)$, are
invariants and all other invariants, included $P(x,q,t)$, derive from them.
It will be useful to classify the non trivial invariants (i.e., not involving
$\psi(q)$ and $V(x,q)$) by the number of derivatives they carry. These
invariants are
\begin{equation}
I_{n_1,n_2}(x)=
\nabla^{n_1}\!\int dk \,k^{n_2}\rho(x,k) = \sum_\alpha p_\alpha
\nabla^{n_1}(P_\alpha\big(\nabla S_\alpha)^{n_2}\big)
\qquad
n_1,n_2=0,1,2,\ldots
\label{eq:4.26}
\end{equation}

Now, the function $P(x,q,t)$ can be computed as a Taylor series in
$t-t_0$. Inspection of the evolution equations (\ref{eq:4.10a}) indicates that
each finite order term in the Taylor expansion will contain a finite number of
derivatives with respect to $x$ and $q$ of the initial data and the
potential. If statistical consistency holds, the combinations of derivatives
allowed cannot be arbitrary, on the contrary they should involve the invariants
in \eq{4.26}. Subsequently we show that this is not the case. The breakdown
occurs for the first time at order $(t-t_0)^4$. The length of the Taylor
coefficients increases rapidly with the order. For this reason, instead of
presenting the proof using generic functions, we take a concrete case which is
sufficient to proof the breakdown of statistical consistency.

Specifically, let the system evolve in $1+1$ dimensions, and
\begin{equation}
t_0=0
,
\quad
P_\alpha(x,q,t_0)= e^{l_\alpha(x)+\kappa q}
,
\quad
S_\alpha(x,q,t_0)=0
,
\quad
V(x,q) = vxq
.
\label{eq:4.28}
\end{equation}
Here $l_\alpha(x)$ are generic functions and $\kappa,v$ are two real
constants.  (The wave function $\psi(q)$ is not normalizable as given. This is
inessential. The proof can be carried out for generic functions, or we can add
a Gaussian factor. Alternatively, the equations are local, and the form
assumed in \eq{4.28} is used only locally.)

The evolution equations (\ref{eq:4.10a}) can be conveniently written in terms
of the variable $L(x,q,t)=\log(P(x,q,t))$,
\begin{equation}
L_t = -\frac{1}{M}(L_x S_x+S_{xx}) -\frac{1}{m}(L_q S_q+S_{qq})
,
\qquad
S_t= -\frac{1}{2M}S_x^2-\frac{1}{2m}S_q^2
+\frac{\hbar^2}{8m}(L_q^2 + 2 L_{qq}) - V
.\label{eq:4.29}
\end{equation}
When $S$ vanishes at $t=0$ and $V$ is an even function of $t$, it follows
(from inspection of \eq{4.29}) that $S$ is an odd function of $t$ and $P$ or
$L$ are even.  (Equivalently, \eq{4.22} admits a solution with $\psi(x,q,t)=
\psi^*(x,q,-t)$.) These conditions hold in our case, so $L_\alpha(x,q,t)$
contains only even powers of $t$ and $S_\alpha(x,q,t)$ contains only odd
powers. Straightforward solution of the equations yields
\begin{eqnarray}
L_\alpha &=& 
l_\alpha + \kappa q
+\left(\frac{v}{M}l_\alpha^\prime q + \frac{v\kappa}{m}x\right)\frac{t^2}{2}
%\\ \nonumber &&
+\left(
-\frac{\hbar^2 \kappa v}{4m M^2}(l_\alpha^\prime l_\alpha^{\prime\prime}
+
l_\alpha^{\prime\prime\prime})
+O(\hbar^0)
\right)\frac{t^4}{4!}
+
O(t^6)
,
\\
\nonumber
S_\alpha &=& 
\left(
\frac{\hbar^2\kappa^2}{8m}-vxq
\right)t
+
\left(
\frac{\hbar^2\kappa v}{4mM}l_\alpha^\prime
-\frac{v^2}{M}q^2-\frac{v^2}{m}x^2
\right)\frac{t^3}{3!}
+
O(t^5)
.
\end{eqnarray}
In the term of order $t^4$ we have omitted contributions without $\hbar$. For
the probability density (using $P_\alpha=e^{L_\alpha}$ in \eq{4.25}) this
implies
\begin{equation}
  \frac{\partial^4 P(x,q,t)}{\partial t^4}\Big|_{t=0}
=
-\frac{\hbar^2 \kappa v}{4m M^2}e^{\kappa q}
\sum_\alpha p_\alpha e^{l_\alpha} (l_\alpha^\prime l_\alpha^{\prime\prime}
+
l_\alpha^{\prime\prime\prime})
+O(\hbar^0)
.
\end{equation}
The sum over $\alpha$ in this expression is not a statistical invariant. From
\eq{4.26}, the invariant with $n_1=3$ and $n_2=0$ is found to be
\begin{equation}
I_{3,0}(x) =
\sum_\alpha p_\alpha e^{l_\alpha} (
(l_\alpha^\prime)^3+
3l_\alpha^\prime l_\alpha^{\prime\prime}
+
l_\alpha^{\prime\prime\prime})
.
\end{equation}
Therefore classical statistical consistency is violated at order $t^4$.  It
can be verified that there is no violation in the purely classical case, that
is, the terms without $\hbar$ involve statistical invariants only, and this is
true also for generic initial data.

\section{Conclusions}
\label{sec:5}

We have discussed the conditions for a perfect quantum-classical hybrid
dynamics and have presented a short and general proof implying that such
perfect hybridization is not viable. This result relies on the assumption that
i) hybrid observables can be obtained by tensor product and that, also in the
hybrid case, there is a bracket which is ii) antisymmetric, iii) a derivation,
and iv) reduces to the standard brackets in each subsector.

We have then introduced a new consistency requirement for quantum-classical
hybrids based on rearrangement invariance of statistical mixtures, that is,
the observable properties of the state should depend on the mixture itself and
not on how it was obtained. Such invariance is automatically fulfilled by
classical and quantum dynamics, but it is a non trivial requirement for
quantum-classical hybrids of non-linear type. In particular, an autonomous
consistent evolution equation for the density matrix cannot be written for the
mean-field approach and so statistical consistency is violated by this scheme.

The statistical ensemble approach in configuration space is analyzed in some
detail. This approach is rather complicated technically since it is highly non
linear. We point out several problematic features of this scheme. First, no
systematic construction of hybrid observables by tensor product is
provided. This avoids the immediate application of the no-go theorem, but it
implies an enormous proliferation of hybrid functionals, the large majority of
which cannot possibly correspond to observables. As a consequence of this
ambiguity it is not known whether the set of observables is closed under the
operation of taking the dynamical bracket.  Certainly one could start by
taking the bracket between classical and quantum observables and then
recursively with the new functionals so generated, to close the minimal
subalgebra containing quantum and classical observables
\cite{Hall:2008}. However, the bracket of a generic classical observable with
a generic quantum one is already so complicated, that our own conjecture is
that such minimal subalgebra is essentially the whole (or a very large) set of
functionals, most of which are certainly not observables. Second, the bracket
of a generic classical observable with a generic quantum observable is not
zero. This leads to ghost interaction between the two sectors when they are
entangled, even after the coupling is no longer present. In particular the
kinetic energy of one sector responds to physical actions on the other sector,
which seems odd. The scheme is designed to treat position-momentum variables,
and so it is of limited applicability in the presence of internal degrees of
freedom, including conservation of spin angular momentum, internal symmetries
and relativistic invariance for particles with spin, such as electrons or
photons.  Statistical consistency of the statistical ensemble approach in
configuration space has been studied too. We find that, as for the mean-field
case, the requirement is not fulfilled, a consequence of the lack of linearity
of the hybrid dynamics. Specifically, it was shown that identical classical
statistical mixtures obtained from different rearrangements evolved
differently after switching on the interaction.

The above analysis tends to reinforce the view that truly classical systems do
not exist in nature, and quantum-classical dynamics are to be regarded as
approximations of fully quantum mechanical systems.

\acknowledgments We thank C. Garc{\'\i}a-Recio for comments along this work.
Research supported by DGI (FIS2008-01143), Junta de Andaluc{\'\i}a grant
FQM-225, the Spanish Consolider-Ingenio 2010 Programme CPAN (CSD2007-00042),
and it is part of the European Community-Research Infrastructure Integrating
Activity Study of Strongly Interacting Matter (acronym HadronPhysics2, Grant
Agreement n. 227431), under the Seventh EU Framework Programme.

%\appendix

%\bibliography{Refs}

\begin{thebibliography}{10}

\bibitem{Borh:1958}
N.~Borh,
\newblock {\em Atomic Physics and Human Knowledge} (Wiley, New York, 1958).
%%CITATION = NONE;%%

\bibitem{Heisenberg:1958}
W.~Heisenberg,
\newblock {\em Physics and Philosophy: The Revolution In Modern Science}
  (Harper Perennial Modern Classics, London, 2007).
%%CITATION = NONE;%%

\bibitem{dEspagnat:1976}
B.~d'Espagnat,
\newblock {\em Conceptual Foundations of Quantum Mechanics} (Addison Wesley,
  Reading, 1976).
%%CITATION = NONE;%%

\bibitem{Allahverdyan:2011cx}
A.~E. Allahverdyan, R.~Balian and T.~M. Nieuwenhuizen,
\newblock 1107.2138.
%%CITATION = ARXIV:1107.2138;%%

\bibitem{DeWitt:1962bu}
B.~S. DeWitt,
\newblock J.Math.Phys. {\bf 3}, 619 (1962).
%%CITATION = JMAPA,3,619;%%

\bibitem{Boucher:1988ua}
W.~Boucher and J.~H. Traschen,
\newblock Phys.Rev. {\bf D37}, 3522 (1988).
%%CITATION = PHRVA,D37,3522;%%

\bibitem{Gisin:1990}
N.~Gisin,
\newblock Phys.Lett. {\bf A143}, 1 (1990).
%%CITATION = NONE;%%

\bibitem{Salcedo:1994sn}
L.~L. Salcedo,
\newblock hep-th/9410007.
%%CITATION = HEP-TH/9410007;%%

\bibitem{Salcedo:1996jr}
L.~L. Salcedo,
\newblock Phys. Rev. {\bf A54}, 3657 (1996), [hep-th/9509089].
%%CITATION = HEP-TH 9509089;%%

\bibitem{Halliwell:1997hy}
J.~Halliwell,
\newblock Phys.Rev. {\bf D57}, 2337 (1998), [quant-ph/9705005].
%%CITATION = QUANT-PH/9705005;%%

\bibitem{Caro:1998us}
J.~Caro and L.~L. Salcedo,
\newblock Phys. Rev. {\bf A60}, 842 (1999), [quant-ph/9812046].
%%CITATION = QUANT-PH 9812046;%%

\bibitem{Peres:2001}
A.~Peres and T.~D. R.,
\newblock Phys. Rev. {\bf A63}, 022101 (2001), [quant-ph/0008068].
%%CITATION = QUANT-PH 0008068;%%

\bibitem{Salcedo:2003mp}
L.~L. Salcedo,
\newblock Phys.Rev.Lett. {\bf 90}, 118901 (2003).
%%CITATION = PRLTA,90,118901;%%

\bibitem{Terno:2004ti}
D.~R. Terno,
\newblock Found.Phys. {\bf 36}, 102 (2006), [quant-ph/0402092].
%%CITATION = QUANT-PH/0402092;%%

\bibitem{Salcedo:2007}
L.~L. Salcedo,
\newblock J. Chem. Phys. {\bf 126}, 057101 (2007).
%%CITATION = JCPSA,126,057101;%%

\bibitem{Agostini:2007}
F.~Agostini, C.~S. and C.~G.,
\newblock Europhys. Lett. {\bf 78}, 30001 (2007).
%%CITATION = NONE;%%

\bibitem{Hu:1994iw}
B.~Hu,
\newblock gr-qc/9403061.
%%CITATION = GR-QC/9403061;%%

\bibitem{Wald:1994bk}
R.~M. Wald,
\newblock {\em Quantum Field Theory in Curved Space-time and Black Hole
  Thermodynamics} (The University of Chicago Press, Chicago, 1994).
%%CITATION = NONE;%%

\bibitem{Craig:2005}
C.~F. Craig, W.~R. Duncan and O.~V. Prezhdo,
\newblock Phys.Rev.Lett. {\bf 95}, 163001 (2005).
%%CITATION = PRLTA,95,163001;%%

\bibitem{Brandenberger:1984cz}
R.~H. Brandenberger,
\newblock Rev.Mod.Phys. {\bf 57}, 1 (1985).
%%CITATION = RMPHA,57,1;%%

\bibitem{Delos:1972}
J.~B. Delos, W.~R. Thorson and S.~K. Knudson,
\newblock Phys.Rev. {\bf A6}, 709 (1972).
%%CITATION = PHRVA,A6,709;%%

\bibitem{Sudarshan:1976bt}
G.~Sudarshan,
\newblock Pramana {\bf 6}, 117 (1976).
%%CITATION = PRAMC,6,117;%%

\bibitem{Sherry:1978ea}
T.~Sherry and E.~Sudarshan,
\newblock Phys.Rev. {\bf D18}, 4580 (1978).
%%CITATION = PHRVA,D18,4580;%%

\bibitem{Aleksandrov:1981}
I.~V. Aleksandrov,
\newblock Z. Naturforsch. {\bf 36A}, 902 (1981).
%%CITATION = ZNASE,36A,902;%%

\bibitem{Jones:1993}
K.~R. Jones,
\newblock Phys.Rev. {\bf A48}, 822–825 (1993).
%%CITATION = PHRVA,A48,822;%%

\bibitem{Anderson:1995tn}
A.~Anderson,
\newblock Phys. Rev. Lett. {\bf 74}, 621 (1995).
%%CITATION = PRLTA,74,621;%%

\bibitem{Diosi:1995qs}
L.~Di\'osi,
\newblock quant-ph/9503023.
%%CITATION = QUANT-PH/9503023;%%

\bibitem{Prezhdo:1996gs}
O.~V. Prezhdo and V.~V. Kisil,
\newblock Phys.Rev. {\bf A56}, 162 (1997), [quant-ph/9610016].
%%CITATION = QUANT-PH/9610016;%%

\bibitem{Diosi:1999py}
L.~Di\'osi, N.~Gisin and W.~T. Strunz,
\newblock Phys.Rev. {\bf A61}, 22108 (2000), [quant-ph/9902069].
%%CITATION = QUANT-PH/9902069;%%

\bibitem{Gindensperger:2000}
E.~Gindensperger, C.~Meier and J.~A. Beswick,
\newblock J. Chem. Phys. {\bf 113}, 9369 (2000).
%%CITATION = JCPSA,113,9369;%%

\bibitem{Dias:2001}
N.~C. Dias,
\newblock J.Math.Phys. {\bf 34}, 771 (2001), [quant-ph/9912071].
%%CITATION = QUANT-PH/9912071;%%

\bibitem{Nielsen:2001wh}
S.~O. Nielsen,
\newblock (2001),
\newblock Ph.D. Thesis (Advisor: Raymond E. Kapral).
%%CITATION = UMI-NQ-63602 ETC.;%%

\bibitem{Prezhdo:2001}
O.~V. Prezhdo and C.~Brooksby,
\newblock Phys. Rev. Lett. {\bf 86}, 3215 (2001).
%%CITATION = PRLTA,86,3215;%%

\bibitem{Kisil:2005}
V.~V. Kisil,
\newblock Europhys. Lett. {\bf 72}, 873 (2005).
%%CITATION = NONE;%%

\bibitem{Sergi:2005}
A.~Sergi,
\newblock Phys.Rev. {\bf E72}, 066125 (2005).
%%CITATION = PHRVA,E72,066125;%%

\bibitem{Prezhdo:2006}
O.~V. Prezhdo,
\newblock J. Chem. Phys. {\bf 124}, 201104 (2006).
%%CITATION = JCPSA,124,201104;%%

\bibitem{Elze:2011hi}
H.-T. Elze,
\newblock 1111.2276.
%%CITATION = ARXIV:1111.2276;%%

\bibitem{Einstein:1935}
A.~Einstein, B.~Podolsky and N.~Rosen,
\newblock Phys.Rev. {\bf 47}, 777 (1935).
%%CITATION = NONE;%%

\bibitem{Bohm:1966}
D.~Bohm and J.~Bub,
\newblock Rev.Mod.Phys. {\bf 38}, 453–469 (1966).
%%CITATION = RMPHA,38,453;%%

\bibitem{Bell:1987hh}
J.~Bell,
\newblock {\em {Speakable and unspeakable in Quantum Mechanics: collected
  papers on quantum philosophy}} (Cambridge University Press, 1987).
%%CITATION = INSPIRE-254492;%%

\bibitem{Schmelzer:2011}
I.~Schmelzer,
\newblock Found.Phys. {\bf 41}, 159 (2011), [0903.4657].
%%CITATION = NONE;%%

\bibitem{Moller:1952}
C.~M{\o}ller,
\newblock {\em {The theory of relativity}} (Clarendon, Oxford, 1952).
%%CITATION = NONE;%%

\bibitem{Rosenfeld:1963}
L.~Rosenfeld,
\newblock Nucl. Phys. {\bf 40}, 353 (1963).
%%CITATION = NUPHA,40,353;%%

\bibitem{Kibble:1979jn}
T.~Kibble and S.~Randjbar-Daemi,
\newblock J.Phys.A {\bf A13}, 141 (1980).
%%CITATION = JPAGB,A13,141;%%

\bibitem{Kibble:1980ia}
T.~Kibble,
\newblock (1980).
%%CITATION = INSPIRE-161985;%%

\bibitem{Isham:1980sb}
C.~Isham,
\newblock (1980).
%%CITATION = ICTP-79-80-36 ETC.;%%

\bibitem{Page:1981aj}
D.~N. Page and C.~Geilker,
\newblock Phys.Rev.Lett. {\bf 47}, 979 (1981).
%%CITATION = PRLTA,47,979;%%

\bibitem{Alvarez:1988tb}
E.~Alvarez,
\newblock Rev.Mod.Phys. {\bf 61}, 561 (1989).
%%CITATION = RMPHA,61,561;%%

\bibitem{Hu:2002jm}
B.~Hu and E.~Verdaguer,
\newblock Class.Quant.Grav. {\bf 20}, R1 (2003), [gr-qc/0211090].
%%CITATION = GR-QC/0211090;%%

\bibitem{Boughn:2008jx}
S.~Boughn,
\newblock Found.Phys. {\bf 39}, 331 (2009), [0809.4218].
%%CITATION = ARXIV:0809.4218;%%

\bibitem{Carlip:2008zf}
S.~Carlip,
\newblock Class.Quant.Grav. {\bf 25}, 154010 (2008), [0803.3456].
%%CITATION = ARXIV:0803.3456;%%

\bibitem{Mattingly:2009}
J.~Mattingly,
\newblock Erkenntnis {\bf 70}, 379 (2009).
%%CITATION = NONE;%%

\bibitem{Diosi:2011vu}
L.~Di\'osi,
\newblock J.Phys.Conf.Ser. {\bf 306}, 012006 (2011), [1101.0672].
%%CITATION = ARXIV:1101.0672;%%

\bibitem{Groenewold:1946}
H.~Groenewold,
\newblock Physica {\bf 12}, 405 (1946).
%%CITATION = NONE;%%

\bibitem{VanHove:1951}
L.~Van~Hove,
\newblock Mem. Acad. R. Belg. {\bf 26}, 61 (1951).
%%CITATION = NONE;%%

\bibitem{Abraham:1978bk}
R.~Abraham and J.~Marsden,
\newblock {\em Foundations of Mechanics} (Benajmin, New York, 1978).
%%CITATION = NONE;%%

\bibitem{Delos:1981}
J.~B. Delos,
\newblock Rev. Mod. Phys. {\bf 53}, 287 (1981).
%%CITATION = RMPHA,53,287;%%

\bibitem{Halcomb:1986}
L.~L. Halcomb and D.~J. Diestler,
\newblock J. Chem. Phys. {\bf 84}, 3130 (1986).
%%CITATION = JCPSA,84,3130;%%

\bibitem{Micha:2007}
D.~A. Micha and I.~Burghardt~(Eds.),
\newblock {\em {Quantum Dynamics of Complex Molecular Systems}} (Springer,
  Berlin, 2007).
%%CITATION = NONE;%%

\bibitem{Bousquet:2011}
D.~Bousquet, K.~H. Hughes, D.~A. Micha and I.~Burghardt,
\newblock J. Chem. Phys. {\bf 134}, 064116 (2011).
%%CITATION = JCPSA,134,064116;%%

\bibitem{Sahoo:2004}
D.~Sahoo,
\newblock J. Phys. {\bf A37}, 997 (2004).
%%CITATION = JPAGB,A37,997;%%

\bibitem{Dass:2009qu}
T.~Dass,
\newblock 0909.4606.
%%CITATION = ARXIV:0909.4606;%%

\bibitem{Carroll:2010zza}
R.~Carroll,
\newblock (2010).
%%CITATION = INSPIRE-855900;%%

\bibitem{Hall:2005ax}
M.~J. Hall and M.~Reginatto,
\newblock Phys.Rev. {\bf A72}, 062109 (2005), [quant-ph/0509134].
%%CITATION = QUANT-PH/0509134;%%

\bibitem{Hall:2008}
M.~J. Hall,
\newblock Phys.Rev. {\bf A78}, 042104 (2008), [0804.2505].
%%CITATION = ARXIV:0804.2505;%%

\bibitem{Reginatto:2009}
M.~Reginatto and M.~J. Hall,
\newblock J.Phys.Conf.Ser. {\bf 174}, 012038 (2009), [0905.2948].
%%CITATION = QUANT-PH/0905.2948;%%

\bibitem{Chua:2011fz}
A.~J. Chua, M.~J. Hall and C.~Savage,
\newblock 1109.3925.
%%CITATION = ARXIV:1109.3925;%%

\bibitem{Diosi:1995fe}
L.~Di\'osi,
\newblock Phys. Rev. Lett. {\bf 76}, 4088 (1996), [quant-ph/9503003].
%%CITATION = QUANT-PH/9503003;%%

\bibitem{Wigner:1932eb}
E.~P. Wigner,
\newblock Phys.Rev. {\bf 40}, 749 (1932).
%%CITATION = PHRVA,40,749;%%

\bibitem{Moyal:1949sk}
J.~Moyal,
\newblock Proc.Cambridge Phil.Soc. {\bf 45}, 99 (1949).
%%CITATION = PCPSA,45,99;%%

\bibitem{Carruthers:1982fa}
P.~Carruthers and F.~Zachariasen,
\newblock Rev.Mod.Phys. {\bf 55}, 245 (1983).
%%CITATION = RMPHA,55,245;%%

\bibitem{Jones:1996}
K.~R. Jones,
\newblock Phys.Rev.Lett. {\bf 76}, 4087 (1996).
%%CITATION = PRLTA,76,4087;%%

\bibitem{Ghirardi:1986}
G.~C. Ghirardi, A.~Rimini and T.~Weber,
\newblock Phys.Rev. {\bf D34}, 470–491 (1986).
%%CITATION = PHRVA,D34,470;%%

\bibitem{Galindo:1991bk}
A.~Galindo and P.~Pascual,
\newblock {\em Quantum Mechanics} (Springer-Verlag, New York, 1991).
%%CITATION = NONE;%%

\bibitem{vonNeumann:1927}
J.~von Neumann,
\newblock {G\"ottinger Nachrichten} {\bf 1}, 245 (1927).
%%CITATION = NONE;%%

\bibitem{dEspagnat:1995}
B.~d'Espagnat,
\newblock {\em Veiled Reality} (Addison Wesley, Reading, 1995).
%%CITATION = NONE;%%

\end{thebibliography}
%\bibliographystyle{JHEP}
%\bibliographystyle{h-physrev4}
%\bibliographystyle{h-elsevier3}

\end{document}